\documentclass[aps,prl,showpacs,reprint]{revtex4-1}

\usepackage{txfonts} 

\usepackage{graphicx}
\usepackage{dcolumn}
\usepackage{bm}
\usepackage{color}
\usepackage{subfigure}
\usepackage{multirow}
\usepackage{amssymb}
\usepackage{hyperref}
\usepackage{eucal}

\newcommand{\iso}[2]{$^{#1}$#2}

\begin{document}
\title{A Novel Source of Tagged Low-Energy Nuclear Recoils}

\author{	Tenzing H.Y. Joshi	}
\affiliation{Department of Nuclear Engineering, University of California, Berkeley, CA 94720, USA} 
\affiliation{Lawrence Livermore National Laboratory, Livermore, CA 94550, USA}

\date{25 May 2011}

\begin{abstract}

For sufficiently wide resonances, nuclear resonance fluorescence behaves like elastic photo-nuclear scattering while retaining the large cross-section characteristic of resonant photo-nuclear absorption. We show that NRF may be used to characterize the signals produced by low-energy nuclear recoils by serving as a novel source of tagged low-energy nuclear recoils. Understanding these signals is important in determining the sensitivity of direct WIMP dark-matter and coherent neutrino-nucleus scattering searches.

\end{abstract}
\maketitle
Direct WIMP dark matter \cite{Gaitskell} and coherent neutrino-nucleus scatter (CNS) \cite{Drukier,Freedman} searches attempt to detect WIMPs and neutrinos respectively by measuring the energy they deposit when scattering off nuclei in a detector. Understanding the way recoiling nuclei distribute energy between scintillation, ionization, and thermal motion as a function of recoil energy is required to define the sensitivity of these rare event detectors. For this purpose the scintillation efficiency ($L_{eff}$) and ionization yield (I.Y.) of nuclear recoils have been reported in candidate detector materials (see Table~\ref{tab:Measurements}). Of the reported measurements, only two \cite{Jones,Barbeau} have probed recoils below one keV, the energy range is the most important to CNS searches \cite{Drukier,Hagmann,Akimov}. Moreover, though the energy domain below a few keV is currently inaccessible in most current WIMP searches, an improved understanding of detector sensitivity in this domain could have significant benefits for setting exclusion limits on WIMPs \cite{Xenon100,Collar}.
\begin{table}[b]
\caption{\small{Previously reported measurements of the relative scintillation efficiency ($L_{eff}$) and ionization yield (I.Y.) from low-energy nuclear recoils in candidate detector materials.}}
\begin{center}
\begin{tabular}{ccccc}
\hline \hline
$ \begin{array}{c} \mbox{Target } \\ \mbox{medium} \end{array} $ 
& $ \begin{array}{c} \mbox{Nuclear recoil} \\ \mbox{energy (keV)} \end{array} $ 
& $ \begin{array}{c} \mbox{Recoil} \\ \mbox{mechanism} \end{array} $ 
& $ \begin{array}{c} \mbox{Measured} \\ \mbox{quantity} \end{array} $ 
& $Ref.$ \\
\hline
LXe & 2--115 & (n,n) & $L_{eff}$ \& I.Y. & \cite{Aprile,Manzur,Sorensen} \\
\cline{1-1}
LAr &11--239 & (n,n) & $L_{eff}$ & \cite{Gastler} \\
\cline{1-1}
\multirow{3}{*}{Ge} & 1--100 & (n,$\text{n}^\prime$) & I.Y. & \cite{Chasman1, Chasman2,Jones2} \\
 & 0.3--1.4 & (n,n) & I.Y. & \cite{Barbeau} \\
 & 0.254 & (n,$\gamma$) & I.Y. & \cite{Jones} \\
\hline \hline
\end{tabular}
\end{center}
\label{tab:Measurements}
\end{table}

It is therefore desirable to experimentally measure the signals produced by nuclear recoils at energies below a few keV. The lowest recoil energy reported \cite{Jones} made use of the nuclear recoil accompanying an (n,$\gamma$) interaction. This approach is hampered by the limited choice of $\gamma$-emitting transitions in the target material and is further complicated by additional energy deposition from secondary $\gamma$-emission. Published measurements at higher recoil energies have employed elastic and inelastic neutron scattering for nuclear recoil production. Those using the latter exploited several low-lying states in Ge isotopes and measured the broadening of their relaxation resulting from the (n,$\text{n}^\prime$) nuclear recoil for varying incident neutron energies \cite{Chasman1,Chasman2,Jones2}. The measurements producing recoils with elastic neutron scattering may be split into two families; the first using mono-energetic neutrons incident on a target and detecting neutrons that scatter in the target at a known scattering angle, the second employ broad spectrum neutron sources and compare data with Monte Carlo simulations. For the purpose of exploring the domain below a few keV, using a mono-energetic neutron source and explicitly tagging the scattered neutron to identify recoil energy is preferable to minimize systematic uncertainty in the measurement. Extending measurements of this type to lower recoil energies may be accomplished by either decreasing the incident neutron energy \cite{Barbeau}, in which case prompt tagging of the scattered neutron becomes a challenge, or decreasing the neutron scattering angle. Both cases pose significant experimental challenges.

In this paper, an alternative approach is presented by considering elastic photo-nuclear scattering rather than elastic neutron-nucleus scattering. As in the neutron experiments the scattered photon may be detected at a known scattering angle, but with the added benefit of energy discrimination in the $\gamma$-tagging detectors, at energies above radioactive backgrounds. The experimental viability of elastic photo-nuclear scattering, both nuclear Thomson and Delbr\"{u}ck, is limited by small cross-sections.  The resonant photo-nuclear scattering process of Nuclear Resonance Fluorsescnce (NRF) does not suffer this limitation. The large cross-section and relatively large width (short lifetime) of some NRF states enable nuclear recoils in the energy domain below a few keV to become accessible to detailed study by providing a novel source of tagged low-energy nuclear recoils.

NRF describes the resonant absorption of a photon by a nucleus and the subsequent relaxation of the excited nucleus via $\gamma$-emission. A detailed discussion of NRF can be found in \cite{Metzger}. The excited nucleus may decay through various allowed channels ($\Gamma_{\text{i}}$), but the branch ($\Gamma_{\text{0}}$) to the ground state will be the focus of this paper. Following photo-absorption, the excited nucleus recoils with momentum equal to that of the incident photon. Governed by the width ($\Gamma$) of the nuclear level, the excited nucleus exists for a finite lifetime ($\tau =\hbar /\Gamma$) before relaxing, in this case via emission of a `fluoresced' photon, again imparting recoil momentum to the nucleus.

\begin{figure}[tb]
\includegraphics[angle=0,width=7.5cm]{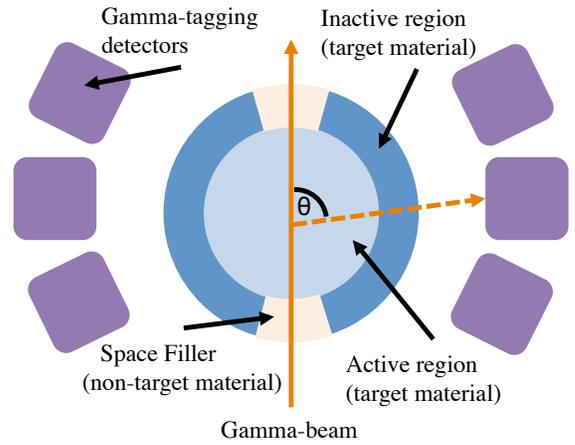}
\caption{\small{(color online) Proposed experimental setup illustrating a $\gamma$-beam incident on a detector with active and inactive regions of a target material. A space-filling material is placed in the beam path to prevent attenuation of resonant photons in the inactive detector region.  The detector is surrounded by an array of $\gamma$-detectors for tagging NRF photons. Image proportions are not to scale.}}
\label{fig:ExpSetup}
\end{figure}

Approximating the energy of the incident and fluoresced photons as the resonance energy ($E_{\text{r}}$), the momentum transferred during NRF is simply $2 E_{\text{r}}\sin(\theta/2)/c$, where $\theta$ is the angle of fluorescence, relative to the direction of the incident photon, in the laboratory frame (see Fig.~\ref{fig:ExpSetup}). Expressing this quantity in terms of kinetic energy, the energy of the nuclear recoil ($E_{\text{NR}}$) resulting from NRF can be described in terms of the resonance energy ($E_{\text{r}}$), mass of the target nucleus ($M$), and the angle of fluorescence ($\theta$).
\begin{equation}
\label{eq:recoilE}
E_{\text{NR}}=\frac{2\left(E_{\text{r}} \sin(\theta /2)\right)^2}{Mc^2}
\end{equation}

This description of final nuclear recoil energy is valid when the lifetime of the excited nuclear state is short enough that no momentum is transferred to the surrounding medium by the recoiling nucleus before photon emission. If this assumption is not met, the momentum transferred during NRF is shared by several atoms, introducing scattering before fluorescence (SBF) to the measurement. The fluoresced photon may still trigger data acquisition. The probability an NRF interaction will result in SBF, or FBS (fluorescence before scattering), is governed by the width of the NRF state, the atomic environment of the target material, the mass of the target nucleus, and the resonance energy. If the mean free path of a nuclear recoil in a given medium is $\ell$, the probability that the recoiling excited nucleus will relax before scattering ($P_{\text{FBS}}$) and thus create a single nuclear recoil with energy related to the angle of fluorescence is given by Eq.~(\ref{eq:FBS}). 
\begin{equation}
\label{eq:FBS}
P_{\text{FBS}}=\ell^{-1}\int _0^{\infty }e^{-x/\ell}\left(1-e^{-x/L}\right)dx
\end{equation}
\noindent The mean distance a recoiling excited nucleus travels before fluorescence is described by $L=E_{\text{r}}\cdot\hbar c /\left(\Gamma \cdot Mc^2\right)$. In a simple fluid the mean free path may be approximated as $\ell = (\text{n$\sigma $})^{-1}$ where n is the local number density and $\sigma$ is the scattering cross-section. If the system is of greater complexity molecular dynamics simulations may be employed to study the probability of fluorescence before scattering. 

The general design of a measurement utilizing NRF to produce nuclear recoils will consist of a photon beam incident on a detector sensitive to the signals produced by a nuclear recoil (see Fig.~\ref{fig:ExpSetup}). The incident $\gamma$-beam produces nuclear recoils in the target region of the detector via NRF, and energy resolving $\gamma$-detectors, placed at different angles of fluorescence, trigger data acquisition upon detection of resonance energy photons. Optimization of an experimental design must take into account selection of target resonances, characteristics of the $\gamma$-beam, geometry and response of the detector, and the $\gamma$-tagging detectors in order to maximize data collection rates. 

A suitable resonance must be selected in an abundant isotope of the target material of the detector by using Eqs.~(\ref{eq:recoilE},\ref{eq:FBS}) and verifying it has a large branching ratio to the ground state. Using only $\gamma$'s emitted during relaxation to the ground state is important in reducing background for two reasons. If a branch to an excited state is used, the energy of the fluoresced $\gamma$ is below beam energy, and can be mimicked by photons created by background processes, resulting in false triggers. Additionally, the recoiling nucleus is still unstable and will likely decay quickly via $\gamma$-emission, depositing additional recoil momentum in the target medium. The density of neighboring NRF states also impacts state selection. Assuming the $\gamma$-tagging detectors are unable to resolve the peaks from neighboring resonances, the presence of several wide states within the envelope of the $\gamma$-source spectrum will increase the event rate at the expense of recoil energy resolution.

Selection of a suitable $\gamma$-source is paramount for the proposed measurement. This application of NRF requires the $\gamma$-source to provide a well collimated beam of photons, of which a significant number are on resonance. For this reason, bremsstrahlung sources are un-usable due to their broad spectrum. A line source from the relaxation of a nuclear level would be ideal, however, identification of a usable line source may pose a challenge. One flexible option is a Compton-backscatter $\gamma$-source at a free electron laser. These facilities are capable of providing tunable pulsed quasi-monoenergetic gamma-beams with energy resolution of one percent \cite{Weller}. 

Beam photons, resonant and nonresonant, will pass through the inactive and active regions of the detector. The mean free path of resonant photons in the target medium will be short due to the large cross-section for resonant absorption. It is therefore important to design or modify a detector in a way that prevents the beam from encountering the target isotope until reaching the active region of the detector to prevent attenuation of resonant photons, like the `space filler' in Fig.~\ref{fig:ExpSetup}. 
The size of the detector's active region may be optimized by minimizing the ratio of statistical uncertainty to event rate considering the characteristics of the $\gamma$-source and detector. 

Non-resonant $\gamma$'s either pass through the detector or interact via photoelectric, Compton scatter, or pair-production. When one of these interactions occur, the detector response will be significantly larger than those produced by the nuclear recoils from NRF, except in the rare case of very low-energy transfer Compton scattering. The large signals make these background events rejectable, a task further simplified in detectors capable of discriminating electronic recoils from nuclear recoils \cite{Lippincott,Manzur}. Though the vast majority of source-related background interactions may be easily rejected, their pileup with NRF events, within the response time of the detector, becomes an irreducible background and reduces rate of usable events. In addition to $P_{\text{FBS}}$, the response times of the detector are then important in constraining the tolerable flux from the $\gamma$-source to maximize the rate of `clean' NRF events. 

Following NRF, detection of the full energy deposition of a fluoresced $\gamma$ serves as the trigger for data acquisition and the tag of nuclear recoil energy. In order to improve $\gamma$-tagging efficiency, an array of $\gamma$ detectors is placed around the target detector, each placed at different fluorescence angles and therefore tagging different recoil energies. The $\gamma$-detectors must be placed outside of the shallow angles where Compton scattered photons are still high in energy and could produce false triggers. The angular domain available for $\gamma$-detector placement is thus constrained by the resolution of the $\gamma$ detectors and the beam energy. The angular position of the $\gamma$ detectors will also impact the expected count rates due to the anisotropy of NRF, which is governed by the initial and final nuclear states as well as the polarization of the incident $\gamma$-beam \cite{Weller}.

Selection of the type of $\gamma$ detectors is important for data collection rates as the overall $\gamma$-tagging efficiency is limited by solid angle and full energy deposition efficiency. Key properties for consideration are $\gamma$ detector material, volume, and response time. Only moderate energy resolution is required because no background process may create photons with energy near the beam energy, $E_{\text{r}}$, assuming the beam/resonance energy is above natural background. Nuclear Thompson, Delbr\"{u}ck, and Rayleigh scattering produce $E_{\text{r}}$ energy photons with very small probabilities, but these processes also produce the nuclear recoil of interest and would thus add to the signal rate. The $\gamma$ detector response time should also influence selection because pileup of background photons in the tagging detectors will increase the dead time in the system. The availability of large crystals with high density, moderate energy resolution, and fast scintillation times make inorganic scintillators strong candidates for this task. 

In some situations, measurements may be performed without $\gamma$-tagging detectors. If beam-related background interactions can be rejected with high efficiency then two types of measurements become available. The largest non-background signals may be attributed to the largest nuclear recoil energy produced by the selected fluorescence state in an end-point measurement. Going one step further, the theoretical distribution of nuclear recoil energies produced by the target resonance could be folded into analysis of the non-background signals in a similar manner to broad-spectrum neutron scatter measurements \cite{Sorensen}. High efficiency background rejection would require detailed background characterization with beam on- and off-resonance in addition to selective triggering of the DAQ and post-processing cuts on quantities such as energy, position, and event shape. 
\begin{figure}[b]
\includegraphics[angle=0,width=8.6cm]{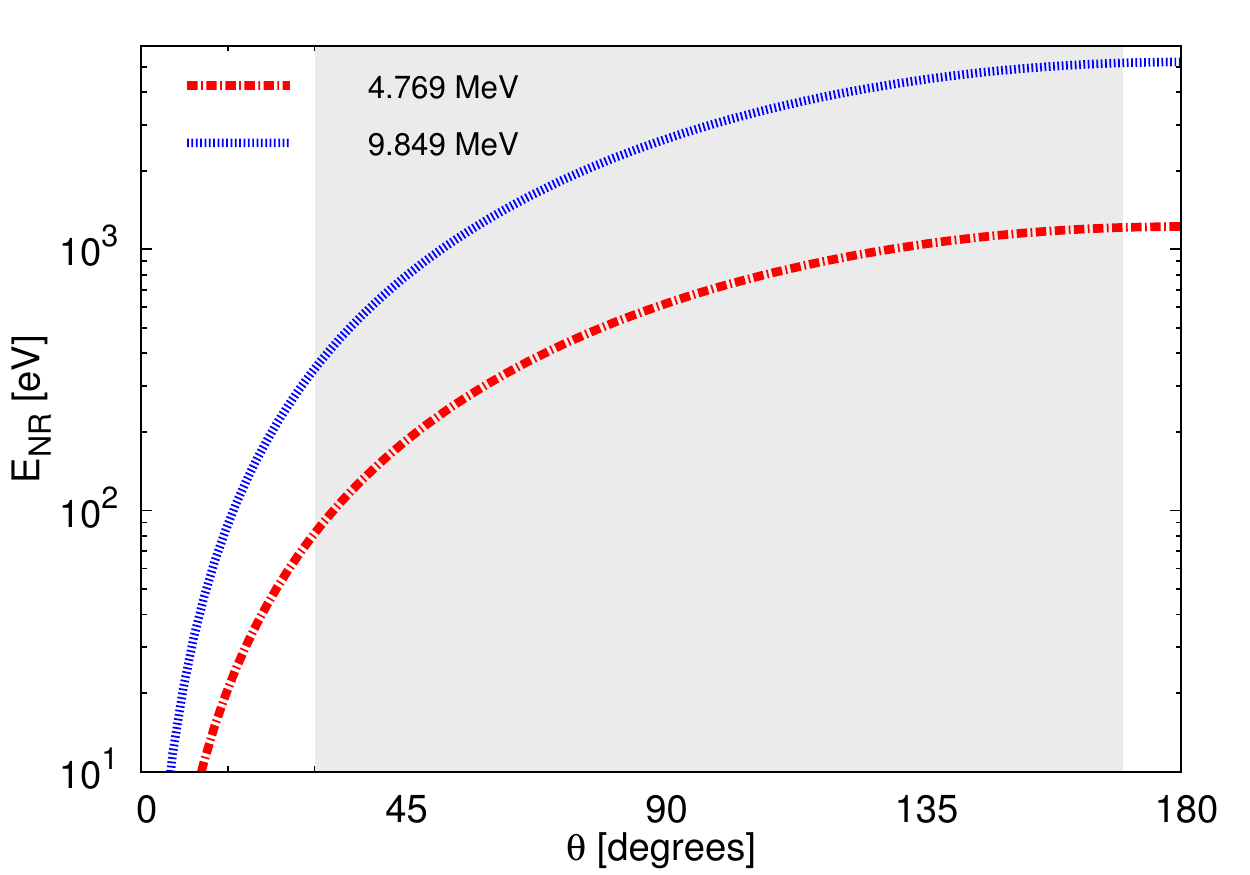}
\caption{\small{(color online) Nuclear recoil energy ($E_{\text{NR}}$), Eq.~\ref{eq:recoilE}, as a function of fluorescence angle ($\theta$) for two candidate resonances in \iso{40}{Ar}. The shaded region shows the angular domain available for $\gamma$-tagging.}}
\label{fig:RecoilSpectrum}
\end{figure}
\begin{table}[t]
\caption{\small{Properties of candidate NRF states in \iso{40}{Ar} \cite{Moreh}. }}
\begin{center}
\begin{tabular}{ccccc}
\hline \hline
$ \begin{array}{c} \mbox{$E_{\text{r}}$ } \\ \mbox{(MeV)} \end{array} $ 
& $ \begin{array}{c} \mbox{$\Gamma$} \\ \mbox{(eV)} \end{array} $ 
& $ \begin{array}{c} \mbox{$\Gamma_{\text{0}}$} \\ \mbox{(eV)} \end{array} $ 
& $P_{\text{FBS}}$ 
& $ \begin{array}{c} \mbox{$E_{\text{NR}}(\theta=\pi)$} \\ \mbox{(keV)} \end{array} $ \\
\hline
4.769 & 0.8 & 0.8 & 0.931 & 1.22 \\
9.356 & 1.0 & 0.57 & 0.896 & 4.70 \\
9.416 & 3.4 & 1.2 & 0.967 & 4.76 \\
9.500 & 7.9 & 6.0 & 0.983 & 4.85 \\
9.582 & 7.3 & 1.6 & 0.984 & 4.93 \\
9.849 & 20.8 & 9.7 & 0.994 & 5.21 \\
9.950 & 9.7 & 1.75 & 0.987 & 5.32 \\
\hline \hline
\end{tabular}
\end{center}
\label{tab:ArRes}
\end{table}

Liquid Argon (LAr) is an ideal case to consider for this novel application of NRF. It is the detector material used in, and proposed for, several dark matter and CNS searches. The level structure of the most abundant isotope, \iso{40}{Ar} (99.6 \%), is well documented up to 10 MeV \cite{Moreh}. Examination of the \iso{40}{Ar} level structure reveals several wide $1^{-}$ levels (E1 transitions) in \iso{40}{Ar}, shown in Table~\ref{tab:ArRes}. The 4.769 MeV state is well suited for production of sub-keV nuclear recoils and the resonances above 9 MeV offer the ability to produce nuclear recoils up to 5.3 keV. Selection between these higher energy states is based on ground state width and isolation. Using these criteria, the 9.849 MeV state is preferable. Fig.~\ref{fig:RecoilSpectrum} shows nuclear recoil energy vs fluorescence angle for the 4.769 and 9.849 MeV states. 

To illustrate the experimental feasibility of this technique we consider a LAr detector with a four cm diameter active region with a 1-cm thick inactive region, filled with PTFE in the beam path to prevent attenuation of the resonant photons (shown in Fig.~\ref{fig:ExpSetup}). Table~\ref{tab:LArRates} gives calculated interaction rates if a pulsed $\gamma$-source producing $10^5$ $\gamma/\text{sec}$ with a one percent energy resolution, characteristic of the High Intensity Gamma-ray Source in high resolution mode (2.79 MHz spill frequency) \cite{Weller}, were incident on this detector. The rate of NRF events ($\dot{R}$) is numerically calculated using the Doppler broadened resonance cross-section \cite{Metzger}. The incident flux and spill frequency of the $\gamma$-source are used to calculate $\Delta$, the fraction of NRF events not lost due to pileup with background interactions Eq.~(\ref{eq:Pnopile}). The total trigger (T) and `clean' trigger (C) rates are given by Eqs.~(\ref{eq:TrigRate}) and (\ref{eq:CleanTrigRate}) respectively. 
\begin{eqnarray}
&& \Delta=\left(\sum _{k=0}^{\infty } \frac{\lambda ^ke^{-\lambda }}{k!} \left(\zeta\right){}^k\right)^x \label{eq:Pnopile} \\
&& T=\eta _{\gamma}\cdot \dot{R} \label{eq:TrigRate} \\
&& C=T \cdot P_{\text{FBS}}\cdot \Delta \label{eq:CleanTrigRate}
\end{eqnarray}
\noindent The probability that a beam-energy photon will pass through the active region of the detector without interaction is $\zeta$, $\lambda$ is the average number of photons expected per spill from the $\gamma$-source ($\lambda = 10^5~\gamma/\text{sec} \times (2.79~\text{MHz})^{-1}$), $k$ is the number of photons per spill, $x$ is the number of spills in two detector response times (assumed to be $2\times30~\mu \text{s}$, $\therefore$ $x=2.79~\text{MHz}\times60~\mu \text{s}=168$), and the full-energy deposition $\gamma$-tagging efficiency is $\eta_\gamma$.
\begin{table}[h]
\caption{\small{Summary of estimated event rates and parameters for two resonance energies ($E_{\text{r}}$). The rate of NRF interactions in the target region is $\dot{R}$ and $\Delta$ describes the fraction of NRF events that are not lost due to pileup with background interactions.  Trigger rates, total (T) and `clean' (C), are given per $\gamma$-tagging detector with full-energy deposition efficiency $\eta_{\gamma}$ at $\theta=\pi/2$.}}
\begin{center}
\begin{tabular}{cccccc}
\hline \hline
$ \begin{array}{c} \mbox{$E_{\text{r}}$} \\ \mbox{(MeV)} \end{array} $ 
& $ \begin{array}{c} \mbox{$\dot{R}$} \\ \mbox{(Hz)} \end{array} $ 
& $ \begin{array}{c} \Delta \end{array} $ 
& $ \eta_{\gamma}$ 
& $ \begin{array}{c} \mbox{T} \\ \mbox{($\text{hour}^{-1}$)} \end{array} $ 
& $ \begin{array}{c} \mbox{C} \\ \mbox{ ($\text{hour}^{-1}$)} \end{array} $ \\
\hline
4.769 & 22.9 & 0.355 & 0.113\% & 94 & 33 \\
9.849 & 49.7 & 0.415 & 0.076\% & 135 & 56 \\
\hline \hline
\end{tabular}
\end{center}
\label{tab:LArRates}
\end{table}

Placement of $\gamma$-tagging detectors at small $\theta$ is limited by the possibility of Compton scattered photons producing false triggers. The finite size of an actual detector smears the limiting tolerable angle. Conservative placement of $\gamma$-tagging detectors at scattering angles larger than 30 degrees prevents Compton-scattered photons and Bremsstrahlung from producing false triggers. Photons elastically scattered through Nuclear Thomson, Delbr\"{u}ck, or Rayleigh interactions by the inactive region of the detector can also produce false triggers. The differential cross-sections of these interactions at relevant angles ($\theta>10~\text{degrees}$) are small, $\mathcal{O} (\mu \text{b/sr})$, and thus their contribution is not included in this exercise.

The estimated event rates suggest that a statistically significant ensemble of tagged energy nuclear recoil events could be collected in several hours of beam-time while operating with an array of $\gamma$-detectors. It is therefore feasible to use NRF to produce nuclear recoils from 0.1--5.2 keV in LAr to characterize scintillation and ionization yields. Similar measurements may also be possible in other materials such as Ge, LXe, and LNe if appropriate resonances can be identified. 

I would like to thank Eric Norman, Adam Bernstein, Chris Angell, and Kareem Kazkaz for their advice and insight and the U.S. Department of Homeland Security's ARI program.  This work was performed under the auspices of the U.S. Department of Energy by Lawrence Livermore National Laboratory under Contract DE-AC52-07NA27344.

\begin{thebibliography}{20}%
\makeatletter
\providecommand \@ifxundefined [1]{%
 \@ifx{#1\undefined}
}%
\providecommand \@ifnum [1]{%
 \ifnum #1\expandafter \@firstoftwo
 \else \expandafter \@secondoftwo
 \fi
}%
\providecommand \@ifx [1]{%
 \ifx #1\expandafter \@firstoftwo
 \else \expandafter \@secondoftwo
 \fi
}%
\providecommand \natexlab [1]{#1}%
\providecommand \enquote  [1]{``#1''}%
\providecommand \bibnamefont  [1]{#1}%
\providecommand \bibfnamefont [1]{#1}%
\providecommand \citenamefont [1]{#1}%
\providecommand \href@noop [0]{\@secondoftwo}%
\providecommand \href [0]{\begingroup \@sanitize@url \@href}%
\providecommand \@href[1]{\@@startlink{#1}\@@href}%
\providecommand \@@href[1]{\endgroup#1\@@endlink}%
\providecommand \@sanitize@url [0]{\catcode `\\12\catcode `\$12\catcode
  `\&12\catcode `\#12\catcode `\^12\catcode `\_12\catcode `\%12\relax}%
\providecommand \@@startlink[1]{}%
\providecommand \@@endlink[0]{}%
\providecommand \url  [0]{\begingroup\@sanitize@url \@url }%
\providecommand \@url [1]{\endgroup\@href {#1}{\urlprefix }}%
\providecommand \urlprefix  [0]{URL }%
\providecommand \Eprint [0]{\href }%
\@ifxundefined \urlstyle {%
  \providecommand \doi  [0]{\begingroup \@sanitize@url \@doi}%
  \providecommand \@doi [1]{\endgroup \@@startlink {\doibase
  #1}doi:\discretionary {}{}{}#1\@@endlink }%
}{%
  \providecommand \doi  [0]{doi:\discretionary{}{}{}\begingroup
  \urlstyle{rm}\Url }%
}%
\providecommand \doibase [0]{http://dx.doi.org/}%
\providecommand \Doi [0]{\begingroup \@sanitize@url \@Doi }%
\providecommand \@Doi  [1]{\endgroup\@@startlink{\doibase#1}\@@Doi}%
\providecommand \@@Doi [1]{#1\@@endlink}%
\providecommand \selectlanguage [0]{\@gobble}%
\providecommand \bibinfo  [0]{\@secondoftwo}%
\providecommand \bibfield  [0]{\@secondoftwo}%
\providecommand \translation [1]{[#1]}%
\providecommand \BibitemOpen [0]{}%
\providecommand \bibitemStop [0]{}%
\providecommand \bibitemNoStop [0]{.\EOS\space}%
\providecommand \EOS [0]{\spacefactor3000\relax}%
\providecommand \BibitemShut  [1]{\csname bibitem#1\endcsname}%
\bibitem [{\citenamefont {Gaitskell}(2004)}]{Gaitskell}%
  \BibitemOpen
  \bibfield  {author} {\bibinfo {author} {\bibfnamefont {R.}~\bibnamefont
  {Gaitskell}},\ }\Doi {10.1146/annurev.nucl.54.070103.181244} {\bibfield
  {journal} {\bibinfo  {journal} {Annu. Rev. Nucl. Part. Sci.},\ }\textbf
  {\bibinfo {volume} {54}},\ \bibinfo {pages} {315} (\bibinfo {year}
  {2004})}\BibitemShut {NoStop}%
\bibitem [{\citenamefont {Drukier}\ and\ \citenamefont
  {Stodolsky}(1984)}]{Drukier}%
  \BibitemOpen
  \bibfield  {author} {\bibinfo {author} {\bibfnamefont {A.}~\bibnamefont
  {Drukier}}\ and\ \bibinfo {author} {\bibfnamefont {L.}~\bibnamefont
  {Stodolsky}},\ }\Doi {10.1103/PhysRevD.30.2295} {\bibfield  {journal}
  {\bibinfo  {journal} {Phys. Rev. D},\ }\textbf {\bibinfo {volume} {30}},\
  \bibinfo {pages} {2295} (\bibinfo {year} {1984})}\BibitemShut {NoStop}%
\bibitem [{\citenamefont {Freedman}(1974)}]{Freedman}%
  \BibitemOpen
  \bibfield  {author} {\bibinfo {author} {\bibfnamefont {D.~Z.}\ \bibnamefont
  {Freedman}},\ }\Doi {10.1103/PhysRevD.9.1389} {\bibfield  {journal} {\bibinfo
   {journal} {Phys. Rev. D},\ }\textbf {\bibinfo {volume} {9}},\ \bibinfo
  {pages} {1389} (\bibinfo {year} {1974})}\BibitemShut {NoStop}%
\bibitem [{\citenamefont {Aprile}\ \emph {et~al.}(2006)\citenamefont {Aprile}
  \emph {et~al.}}]{Aprile}%
  \BibitemOpen
  \bibfield  {author} {\bibinfo {author} {\bibfnamefont {E.}~\bibnamefont
  {Aprile}} \emph {et~al.},\ }\Doi {10.1103/PhysRevLett.97.081302} {\bibfield
  {journal} {\bibinfo  {journal} {Phys. Rev. Lett.},\ }\textbf {\bibinfo
  {volume} {97}},\ \bibinfo {pages} {081302} (\bibinfo {year}
  {2006})}\BibitemShut {NoStop}%
\bibitem [{\citenamefont {Manzur}\ \emph {et~al.}(2010)\citenamefont {Manzur}
  \emph {et~al.}}]{Manzur}%
  \BibitemOpen
  \bibfield  {author} {\bibinfo {author} {\bibfnamefont {A.}~\bibnamefont
  {Manzur}} \emph {et~al.},\ }\Doi {10.1103/PhysRevC.81.025808} {\bibfield
  {journal} {\bibinfo  {journal} {Phys. Rev. C},\ }\textbf {\bibinfo {volume}
  {81}},\ \bibinfo {pages} {025808} (\bibinfo {year} {2010})}\BibitemShut
  {NoStop}%
\bibitem [{\citenamefont {Sorensen}\ \emph {et~al.}(2009)\citenamefont
  {Sorensen} \emph {et~al.}}]{Sorensen}%
  \BibitemOpen
  \bibfield  {author} {\bibinfo {author} {\bibfnamefont {P.}~\bibnamefont
  {Sorensen}} \emph {et~al.},\ }\href
  {http://www.sciencedirect.com/science/article/B6TJM-4VBDGYK-3/2/955ce6ee62330f437f352d46bbc8f500}
  {\bibfield  {journal} {\bibinfo  {journal} {Nucl. Instr. Meth. A},\ }\textbf
  {\bibinfo {volume} {601}},\ \bibinfo {pages} {339 } (\bibinfo {year}
  {2009})}\BibitemShut {NoStop}%
\bibitem [{\citenamefont {Gastler}\ \emph {et~al.}(2010)\citenamefont {Gastler}
  \emph {et~al.}}]{Gastler}%
  \BibitemOpen
  \bibfield  {author} {\bibinfo {author} {\bibfnamefont {D.}~\bibnamefont
  {Gastler}} \emph {et~al.},\ }\href {http://arxiv.org/pdf/1004.0373} {
  (\bibinfo {year} {2010})},\ \Eprint {http://arxiv.org/abs/1004.0373}
  {arXiv:1004.0373 [physics.ins-det]} \BibitemShut {NoStop}%
\bibitem [{\citenamefont {Chasman}\ \emph {et~al.}(1965)\citenamefont {Chasman}
  \emph {et~al.}}]{Chasman1}%
  \BibitemOpen
  \bibfield  {author} {\bibinfo {author} {\bibfnamefont {C.}~\bibnamefont
  {Chasman}} \emph {et~al.},\ }\Doi {10.1103/PhysRevLett.15.245} {\bibfield
  {journal} {\bibinfo  {journal} {Phys. Rev. Lett.},\ }\textbf {\bibinfo
  {volume} {15}},\ \bibinfo {pages} {245} (\bibinfo {year} {1965})}\BibitemShut
  {NoStop}%
\bibitem [{\citenamefont {Chasman}\ \emph {et~al.}(1967)\citenamefont {Chasman}
  \emph {et~al.}}]{Chasman2}%
  \BibitemOpen
  \bibfield  {author} {\bibinfo {author} {\bibfnamefont {C.}~\bibnamefont
  {Chasman}} \emph {et~al.},\ }\Doi {10.1103/PhysRev.154.239} {\bibfield
  {journal} {\bibinfo  {journal} {Phys. Rev.},\ }\textbf {\bibinfo {volume}
  {154}},\ \bibinfo {pages} {239} (\bibinfo {year} {1967})}\BibitemShut
  {NoStop}%
\bibitem [{\citenamefont {Jones}\ and\ \citenamefont {Kraner}(1971)}]{Jones2}%
  \BibitemOpen
  \bibfield  {author} {\bibinfo {author} {\bibfnamefont {K.~W.}\ \bibnamefont
  {Jones}}\ and\ \bibinfo {author} {\bibfnamefont {H.~W.}\ \bibnamefont
  {Kraner}},\ }\Doi {10.1103/PhysRevC.4.125} {\bibfield  {journal} {\bibinfo
  {journal} {Phys. Rev. C},\ }\textbf {\bibinfo {volume} {4}},\ \bibinfo
  {pages} {125} (\bibinfo {year} {1971})}\BibitemShut {NoStop}%
\bibitem [{\citenamefont {Barbeau}\ \emph {et~al.}(2007)\citenamefont {Barbeau}
  \emph {et~al.}}]{Barbeau}%
  \BibitemOpen
  \bibfield  {author} {\bibinfo {author} {\bibfnamefont {P.~S.}\ \bibnamefont
  {Barbeau}} \emph {et~al.},\ }\href
  {http://stacks.iop.org/1475-7516/2007/i=09/a=009} {\bibfield  {journal}
  {\bibinfo  {journal} {JCAP},\ }\textbf {\bibinfo {volume} {2007}},\ \bibinfo
  {pages} {009} (\bibinfo {year} {2007})}\BibitemShut {NoStop}%
\bibitem [{\citenamefont {Jones}\ and\ \citenamefont {Kraner}(1975)}]{Jones}%
  \BibitemOpen
  \bibfield  {author} {\bibinfo {author} {\bibfnamefont {K.~W.}\ \bibnamefont
  {Jones}}\ and\ \bibinfo {author} {\bibfnamefont {H.~W.}\ \bibnamefont
  {Kraner}},\ }\Doi {10.1103/PhysRevA.11.1347} {\bibfield  {journal} {\bibinfo
  {journal} {Phys. Rev. A},\ }\textbf {\bibinfo {volume} {11}},\ \bibinfo
  {pages} {1347} (\bibinfo {year} {1975})}\BibitemShut {NoStop}%
\bibitem [{\citenamefont {Hagmann}\ and\ \citenamefont
  {Bernstein}(2004)}]{Hagmann}%
  \BibitemOpen
  \bibfield  {author} {\bibinfo {author} {\bibfnamefont {C.}~\bibnamefont
  {Hagmann}}\ and\ \bibinfo {author} {\bibfnamefont {A.}~\bibnamefont
  {Bernstein}},\ }\Doi {10.1109/TNS.2004.836061} {\bibfield  {journal}
  {\bibinfo  {journal} {IEEE Trans. Nucl. Sci.},\ }\textbf {\bibinfo {volume}
  {51}},\ \bibinfo {pages} {2151 } (\bibinfo {year} {2004})},\ ISSN \bibinfo
  {issn} {0018-9499}\BibitemShut {NoStop}%
\bibitem [{\citenamefont {Akimov}\ \emph {et~al.}(2009)\citenamefont {Akimov}
  \emph {et~al.}}]{Akimov}%
  \BibitemOpen
  \bibfield  {author} {\bibinfo {author} {\bibfnamefont {D.}~\bibnamefont
  {Akimov}} \emph {et~al.},\ }\href
  {http://stacks.iop.org/1748-0221/4/i=06/a=P06010} {\bibfield  {journal}
  {\bibinfo  {journal} {JINST},\ }\textbf {\bibinfo {volume} {4}},\ \bibinfo
  {pages} {P06010} (\bibinfo {year} {2009})}\BibitemShut {NoStop}%
\bibitem [{\citenamefont {Aprile}\ \emph {et~al.}(2010)\citenamefont {Aprile}
  \emph {et~al.}}]{Xenon100}%
  \BibitemOpen
  \bibfield  {author} {\bibinfo {author} {\bibfnamefont {E.}~\bibnamefont
  {Aprile}} \emph {et~al.} (\bibinfo {collaboration} {XENON100
  Collaboration}),\ }\Doi {10.1103/PhysRevLett.105.131302} {\bibfield
  {journal} {\bibinfo  {journal} {Phys. Rev. Lett.},\ }\textbf {\bibinfo
  {volume} {105}},\ \bibinfo {pages} {131302} (\bibinfo {year}
  {2010})}\BibitemShut {NoStop}%
\bibitem [{\citenamefont {Collar}\ and\ \citenamefont
  {McKinsey}(2010)}]{Collar}%
  \BibitemOpen
  \bibfield  {author} {\bibinfo {author} {\bibfnamefont {J.~I.}\ \bibnamefont
  {Collar}}\ and\ \bibinfo {author} {\bibfnamefont {D.~N.}\ \bibnamefont
  {McKinsey}},\ }\href@noop {} { (\bibinfo {year} {2010})},\ \Eprint
  {http://arxiv.org/abs/1005.0838} {arXiv:1005.0838 [astro-ph.CO]} \BibitemShut
  {NoStop}%
\bibitem [{\citenamefont {Metzger}(1959)}]{Metzger}%
  \BibitemOpen
  \bibfield  {author} {\bibinfo {author} {\bibfnamefont {F.}~\bibnamefont
  {Metzger}},\ }\enquote {\bibinfo {title} {Resonance fluorescence in
  nuclei},}\ in\ \href@noop {} {\emph {\bibinfo {booktitle} {Progress in
  Nuclear Physics}}},\ Vol.~\bibinfo {volume} {7}\ (\bibinfo {year} {1959})\
  pp.\ \bibinfo {pages} {54--88}\BibitemShut {NoStop}%
\bibitem [{\citenamefont {Weller}\ \emph {et~al.}(2009)\citenamefont {Weller}
  \emph {et~al.}}]{Weller}%
  \BibitemOpen
  \bibfield  {author} {\bibinfo {author} {\bibfnamefont {H.~R.}\ \bibnamefont
  {Weller}} \emph {et~al.},\ }\href
  {http://www.sciencedirect.com/science/article/B6TJC-4T3VR6N-1/2/d647b4cb81fb15b6a0bbee7907657256}
  {\bibfield  {journal} {\bibinfo  {journal} {Prog. Particle and Nucl. Phys.},\
  }\textbf {\bibinfo {volume} {62}},\ \bibinfo {pages} {257 } (\bibinfo {year}
  {2009})},\ ISSN \bibinfo {issn} {0146-6410}\BibitemShut {NoStop}%
\bibitem [{\citenamefont {Lippincott}\ \emph {et~al.}(2008)\citenamefont
  {Lippincott} \emph {et~al.}}]{Lippincott}%
  \BibitemOpen
  \bibfield  {author} {\bibinfo {author} {\bibfnamefont {W.~H.}\ \bibnamefont
  {Lippincott}} \emph {et~al.},\ }\Doi {10.1103/PhysRevC.78.035801} {\bibfield
  {journal} {\bibinfo  {journal} {Phys. Rev. C},\ }\textbf {\bibinfo {volume}
  {78}},\ \bibinfo {pages} {035801} (\bibinfo {year} {2008})}\BibitemShut
  {NoStop}%
\bibitem [{\citenamefont {Moreh}\ \emph {et~al.}(1988)\citenamefont {Moreh}
  \emph {et~al.}}]{Moreh}%
  \BibitemOpen
  \bibfield  {author} {\bibinfo {author} {\bibfnamefont {R.}~\bibnamefont
  {Moreh}} \emph {et~al.},\ }\Doi {10.1103/PhysRevC.37.2418} {\bibfield
  {journal} {\bibinfo  {journal} {Phys. Rev. C},\ }\textbf {\bibinfo {volume}
  {37}},\ \bibinfo {pages} {2418} (\bibinfo {year} {1988})}\BibitemShut
  {NoStop}%
\end{thebibliography}
\end{document}